\documentclass[reprint,aps,amsmath,amssymb,nofootinbib,twoside,prd,showkeys,superscriptaddress]{revtex4-1}
\pdfoutput=1
\usepackage{verbatim}
\usepackage{comment}
\usepackage{graphicx}
\usepackage[T1]{fontenc}
\usepackage{epsfig}
\usepackage{bm}
\usepackage{tensor}
\usepackage{amssymb}
\usepackage{amsmath}
\usepackage{subfigure}
\usepackage{dcolumn}
\usepackage[utf8]{inputenc}
\usepackage{cancel}
\usepackage[colorlinks]{hyperref}
\usepackage[usenames,dvipsnames]{color}
\usepackage{float}
\hypersetup{
     breaklinks=true,
    pdfstartview={FitH},    
    colorlinks=true,       
    linkcolor=blue,          
    citecolor=red,        
    filecolor=magenta,      
    urlcolor=blue,           
    anchorcolor=green,      
    linktocpage=true
}

\usepackage{wrapfig}

\def\doi{http://doi.org}

\newcommand{\aj}{Astron. J.}   
\newcommand{\aap}{Astron. Astrophys.}   
\newcommand{\mnras}{Mon. Not. R. Astron. Soc.}   
\newcommand{\na}{New Astron.}   

\DeclareUnicodeCharacter{2212}{-}

\newcommand{\HCd}{\mathcal{H}}

\def\HCdt0{\tilde{\HCd}_{0}}

\usepackage[normalem]{ulem}  
\newcommand\redsout{\bgroup\markoverwith{\textcolor{red}{\rule[0.5ex]{10pt}{2.pt}}}\ULon}  

\newcommand{\affLib}{Leibniz-Institut fur Astrophysik Potsdam, An der Sternwarte 16, D-14482 Potsdam, Germany} 

\begin{document}
\title{Multiscale Cosmic Curvature: from Local Structures to Cosmology}
\author{David Benisty}
\email{benidav@aip.de}
\affiliation{\affLib}
\begin{abstract}
This study tackles the impact Dark Energy (DE) in different systems by a simple unifying formalism. We introduce a parameter space that compare gravity across all cosmic scales, using the McVittie spacetime (McV) and connects spherically symmetric solutions with cosmological solutions. By analyzing the invariant scalars: the Ricci, Weyl, and Kretschmann scalars, we develop a phase-space description that predicts the dominance of the Cosmological Constant. We explore three cases: (1) the local Hubble flow around galaxy groups and clusters, (2) spherical density distributions and (3) binary motion.  {Our results show that the Kretschmann scalar of galaxy groups and clusters in their turnaround is $2\Lambda^2$ which is three times the Kretschmann scalar of the Cosmological Consonant. This indicates the DE domination in local structures.}
\end{abstract}
\maketitle
\section{Introduction} 
Dark Energy (DE) remains one of the most elusive components of modern cosmology. First introduced to explain the observed accelerated expansion of the universe, DE is thought to be a pervasive form of energy with negative pressure, capable of driving repulsive gravitational effects that counterbalance the attraction of matter on cosmic scales~\cite{Peebles:2002gy}. The discovery of the unexpected dimming of distant Type Ia supernovae provided the first compelling evidence for this component, fundamentally reshaping our understanding of cosmic evolution and the large-scale structure of the Universe~\cite{SupernovaCosmologyProject:1998vns}. Subsequent measurements of the cosmic microwave background (CMB)\cite{Planck:2018vyg}, baryon acoustic oscillations (BAO), and large-scale structure surveys have since reinforced DE’s role as a dominant constituent of the cosmic energy budget\cite{CosmoVerseNetwork:2025alb}.

 {While DE is commonly treated as a smooth background field, its influence on local dynamics—particularly in environments such as galaxy clusters and near zero-velocity surfaces—offers a complementary means of probing its properties~\cite{Pavlidou:2004vq,Nasonova:2011md,Pavlidou:2013zha,Pavlidou:2014aia,Tanoglidis:2014lea,Tanoglidis:2016lrj,Lopes:2018uhq,Silbergleit:2019oyx,Bhattacharya:2019bgg,Kim:2020gai,Pavlidou:2020afx,Paraskevas:2023itu,DelPopolo:2020ihg,DelPopolo:2020mge}. Transitional region where cosmic expansion balances local gravitational attraction—are especially valuable for this purpose, serving as natural laboratories for testing how DE interacts with matter on intermediate scales~\cite{bib:Sandage1986,Peirani:2005ti,Peirani:2008qs,Penarrubia:2014oda,Sorce:2016yok,Kim:2020gai,DelPopolo:2021hkz,DelPopolo:2022sev,Aluri:2022hzs,Wagner:2025wrp}.}

 {A natural theoretical framework for exploring these multiscale effects is the McVittie (McV) spacetime~\cite{McVittie:1933zz}, which describes a spherically symmetric mass embedded in an expanding cosmological background~\cite{Faraoni:2007es,Nolan:1998xs}. The McV metric interpolates between Schwarzschild-like strong-field gravity~\cite{Antoniou:2016obw} and FLRW cosmology, making it well suited for studying the interplay between local gravitational fields and cosmic expansion~\cite{Nandra:2011ug,Nandra:2011ui}. Curvature invariants such as the Ricci scalar, Kretschmann scalar, and Weyl invariants provide powerful tools for diagnosing this interplay~\cite{Iihoshi:2007uz,Baker:2014zba,Moulin:2024yxf}. Their behavior can reveal regions where DE dominates, quantify tidal and shear effects, and trace how local overdensities modulate the effective strength of cosmic acceleration, extending analyses such as~\cite{Baker:2014zba}.}

 {In inhomogeneous regions, deviations from ideal FLRW behavior are encoded in curvature invariants, offering a diagnostic of DE-driven dynamics. For example, the Kretschmann scalar highlights the transition from mass-dominated to DE-dominated regimes, while Weyl curvature encodes the anisotropic imprint of local structures~\cite{Baker:2014zba}. Linking these curvature measures to local kinematics bridges theoretical predictions with observations, from the dynamics of binaries~\cite{Benisty:2023vbz,Benisty:2023clf} and dwarf galaxies in groups and clusters~\cite{Karachentsev:2008st,Karachentsev:2006ww,karachentsev2002,Karachentsev:2013cva,2012AJ....144....4M,2020AJ....160..124M,2020RNAAS...4..229M,Tully:2022rbj} to cluster virialization and the universe’s large-scale expansion~\cite{1972ApJ...176....1G,1980lssu.book.....P,Pace:2017qxv}. Comparisons between predicted DE-dominant regions and observed peculiar velocity fields allow us to test whether local expansion is consistent with a cosmological constant or indicates more complex DE behavior. With the advent of high-precision galaxy kinematic surveys and cosmological simulations~\cite{Nelson:2018uso}, it is now possible to map spatial variations in the local expansion rate and use them as a probe of DE–gravity interplay.}

 {In this work, we study DE phenomena across cosmological and local scales within the McVittie framework. By computing curvature invariants and analyzing geodesic motion, we identify regimes where DE effects prevail over local gravity and assess their observational consequences for local Hubble flows. Our approach connects theoretical diagnostics with empirical data, providing a unified framework to test DE’s influence from compact astrophysical systems to the largest cosmological structures. }

The paper is structured as follows: Section~\ref{sec:spacetime} introduces the McVittie spacetime and its limiting cases. Section~\ref{sec:geo} presents the geodesic equations in this setting. Section~\ref{sec:expan} analyzes the regimes of expansion dominance using different datasets. Section~\ref{sec:dis} concludes with a discussion of results and their implications for future DE studies.

\section{General Spacetime}  
\label{sec:spacetime}
\subsection{The Curvature}
A general framework for describing bound systems embedded in an expanding cosmological background is provided by the McV metric~\cite{McVittie:1933zz,Nandra:2011ug}. For a flat cosmological background, this metric takes the form:  
\begin{eqnarray}
ds^2 = -\left(1 - \Phi - \frac{r^2 H^2}{c^2} \right) c^2 dt^2 - \frac{2 r H}{\sqrt{1 - \Phi}} c \, dt \, dr \\  + \nonumber  \frac{dr^2}{1 - \Phi} + r^2 d\Omega^2,
\label{eq:metric}
\end{eqnarray}
where:
\begin{equation}
\Phi \equiv \frac{2 G M}{r c^2} ,\quad H \equiv \frac{\dot{a}}{a},
\end{equation}
 {represents the gravitational potential $\Phi $ and the Hubble parameter $H$ and $\Omega$ is the angular part of the metric. $M$ is the total mass of the structure, $G$ is the Newtonian gravitational constant and $c$ is the speed of light and $a$ is the scale factor of the spacetime. In Eq. (\ref{eq:metric}), $r$ represents the physical spatial coordinate, which is related to the comoving coordinate by $\chi = r/a(t)$.}

 {This choice of physical coordinates (as opposed to isotropic coordinates) is motivated by its direct physical interpretation for local systems and its straightforward relation to the comoving radius $\chi$, which naturally describes the expanding background in standard FLRW cosmology. The metric reduces to the FLRW metric in the absence of a local mass ($\Phi = 0$ and $M = 0$), and to the Schwarzschild metric when the Hubble parameter vanishes ($H = 0$), corresponding to a static, isolated mass. In the special case where $H = \text{Const}$ and $\Phi = 0$, the metric simplifies to the de Sitter metric, describing a universe dominated by a cosmological constant. }

 {By setting $M = 0$ and using the comoving coordinate, we recover the standard flat FLRW metric, which describes a homogeneous and isotropic universe:}  
\begin{eqnarray}  
ds^2 = - (c^2 - r^2 H^2) \, dt^2 - 2rH \, dt \, dr + dr^2 + r^2 d\Omega^2 \nonumber\\ = -c^2 dt^2 + a^2(d\chi^2 + \chi^2 d\Omega^2).  
\end{eqnarray}  
Similarly, setting $H = 0$ reduces the metric to the Schwarzschild metric~\cite{Antoniou:2016obw}.

 {To investigate how local and cosmological effects combine in this spacetime, it is useful to compute the connection coefficients. For simplicity, we define the parameters:}

\begin{align}
S \equiv \sqrt{1-\Phi}, \quad D_H \equiv \frac{c}{H} \quad w \equiv -1 - \frac{2\dot{H}}{3H^{2}}. 
\end{align}

 {where $D_H$ is the Hubble distance and $w$ is the equation of state of the Universe. For a DE-dominated universe, $w = -1$; for a matter-dominated universe, $w = 0$; and for a radiation-dominated era, $w = 1/3$. Using these abbreviations, the Christoffel symbols can be written in an elegant way:}

\begin{align*}
\Gamma^{t}{}_{tt} &= \frac{2r^{2} - D_H^{2}\Phi}{2D_H^{3}S}, \quad 
\Gamma^{t}{}_{tr} = \frac{\Phi - \tfrac{2r^{2}}{D_H^{2}}}{2r S^2}, \\
\Gamma^{t}{}_{rr} &= \frac{1}{D_H S^3}, \quad
\Gamma^{t}{}_{\theta\theta} = \frac{r^{2}}{D_H S}, \quad \Gamma^{t}{}_{\phi\phi} = \frac{r^{2}\sin^{2}\theta}{D_H S}, \\
\Gamma^{r}{}_{tt} &= \frac{r^{3}}{D_H^{4}} + \frac{r}{2D_H^{2}}(1 - S) - \frac{S^2 \Phi}{2r} + \frac{3r S}{2D_H^2} (1+w), \\
\Gamma^{r}{}_{tr} &= \frac{D_H^{2} \Phi -2r^{2} }{2D_H^{3}S}, \quad
\Gamma^{r}{}_{rr} = \frac{D_H^{2}\Phi -2r^{2} }{2D_H^{2}r S^2}, \\
\Gamma^{r}{}_{\theta\theta} &= r\left(\frac{r^{2}}{D_H^{2}} -S \right), \quad
\Gamma^{r}{}_{\phi\phi} = \frac{r^{2}-D_H^{2}S^2}{D_H^{2}}r\sin^{2}\theta,
\\
\Gamma^{\theta}{}_{r\theta} &= \Gamma^{\phi}{}_{r\phi} = \frac{1}{r}, \quad
\Gamma^{\theta}{}_{\phi\phi} = -\sin\theta\cos\theta, \quad \Gamma^{\phi}{}_{\theta\phi} = \cot\theta.
\end{align*}
 {Together with the radial and angular components, which reproduce the usual Schwarzschild and FLRW limits in the appropriate regimes. From these one constructs the Riemann tensor, whose independent nonzero components are:}

\begin{align*}
R_{\hat t\hat r\hat t\hat r}
&= \frac{1}{D_{H}^{2}}-\frac{\Phi^{2}}{2r^2} 
\;+\;\frac{3}{2}\,(1+w)\,\frac{S}{D_{H}^{2}},
\\[8pt]
R_{\hat t\hat\theta\hat t\hat\theta}
&= R_{\hat r\hat\theta\hat r\hat\theta} = \frac{1}{2}\!\left(\Phi-\frac{2r^{2}}{D_{H}^{2}}\right),
\\[8pt]
R_{\hat r\hat\theta\hat\theta\hat t}
&= \frac{3}{2}\,(1+w)\,\frac{r \, c}{D_{H}^{3}}\;,
\\[8pt]
R_{\hat r\hat\phi\hat\phi\hat t}
&=\;\frac{3}{2}\,(1+w)\,\frac{r\, c}{D_{H}^{3}}\;\,\sin^{2}\theta,
\\[8pt]
R_{\hat\theta\hat\phi\hat\theta\hat\phi}
&= -\Big(\Phi+\frac{r^{2}}{D_{H}^{2}}\Big)\,\sin^{2}\theta .
\end{align*}

 {The above curvature components clearly interpolate between the Schwarzschild tidal terms and the cosmological background curvature. To understand the matter content that sources this geometry, we next evaluate the Einstein tensor $G_{\mu\nu}$. Its components take the form:}

\begin{align}
G^{0}{}_{0} &= \frac{3}{D_H^2}, \quad G^{1}{}_{0} = \frac{3 r c}{D_H^{2}} (1+w), \nonumber
\\
G^{j}{}_{j} &= - \frac{3}{D_H^2}
\left(1 - \frac{1+w}{S}\right),
\end{align}
with $j=1,2,3$ labeling the spatial components.
 {The mixed component encodes a momentum flux proportional to $(1+w)$, vanishing in the cosmological constant case. The spatial components reveal anisotropic stresses induced by the local gravitational potential $\Phi$, signaling a departure from a pure perfect-fluid form in the presence of a compact object. Physically, the energy–momentum tensor thus consists of a homogeneous cosmological fluid driving the expansion, together with an inhomogeneous contribution due to the local mass. The combined source reproduces the hybrid McV geometry, which smoothly interpolates between the FLRW and Schwarzschild limits.}

 {A distinctive feature of this effective matter source is that, unlike the strictly homogeneous FLRW case, the pressure acquires a nontrivial radial dependence through the potential $\Phi$. Close to the surface $r = 2M/a(t)$ this pressure diverges, leading to violations of the strong energy condition \cite{Nolan:1998xs,Faraoni:2007es}. These pathologies underline that the McV spacetime, while exact, is not supported by a physically realistic perfect fluid near the would-be black hole horizon.}

 {The horizon structure itself is also subtle. In dust-dominated backgrounds ($w=0$) the McV horizon evolves dynamically, can fail to form smoothly, and may even bifurcate into multiple branches \cite{Nolan:1998xs,Faraoni:2007es}. This peculiar behavior complicates the interpretation of how local Schwarzschild-like physics transitions into the cosmological FLRW regime. It should be kept in mind that extending the discussion into the near-horizon region would require caution, since violation of the energy condition and the dynamical horizon structure could substantially affect physical conclusions.}

To analyze the curvature properties of the McV spacetime, we compute three key curvature scalars: the Ricci scalar $\mathcal{R}$, the Weyl scalar $\mathcal{C}$, and the Kretschmann scalar $\mathcal{K}$. The Ricci scalar $\mathcal{R}$ reflects the overall curvature influenced by both the local mass and the cosmological expansion. The Weyl scalar $\mathcal{C}$ characterizes the tidal forces generated by the local mass, which dominate near compact objects.  {This scalar depends solely on the local mass because the FLRW background is conformally flat and thus has vanishing Weyl curvature, supported by \cite{Iihoshi:2007uz} in Eq.~(4.28) and \cite{Moulin:2024yxf}.} The Kretschmann scalar $\mathcal{K}$ combines contributions from both local and global curvature, offering a complete measure of spacetime curvature invariants:  
\begin{align}  
\mathcal{R} &=  \frac{12}{D_H^2}  \left(1 - \frac{3(1 + w )}{4S}\right), \quad   
\mathcal{C} = 12 \left(\frac{ \Phi}{r^2} \right)^2, \nonumber \\  
\mathcal{K} &=  {12} \left(\frac{\Phi}{r^2}\right)^2 + \frac{3}{D_H^4} \left(8 -\frac{12 (1 + w)}{S} + \frac{9 (1 + w)^2}{S^2}\right),  
\end{align}  
To quantify the relative strength of local gravitational effects compared to cosmic expansion, we introduce the following  {physically motivated} dimensionless parameters:  
\begin{eqnarray}  
\kappa \equiv \frac{\Phi}{r^2} \bigg/ \frac{2H^2}{c^2} = \frac{G M}{H^2 r^3}, \quad  \tau \equiv \frac{1 + w}{\sqrt{1-\Phi }}.  
\label{eq:curv_scalar}
\end{eqnarray}  
 {$\kappa$ represents the ratio of local density ($\sim M/r^3$) to cosmological critical density ($3H^2/8\pi G$), while $\tau$ encodes the equation-of-state modification by gravitational potential.} The McV metric enables the investigation of how bound systems, such as galaxies or galaxy clusters, evolve within an expanding cosmological background. The dimensionless parameter $\kappa$ determines the ratio between these two parts and as we will see from different systems has different interpretation. 

\section{The Geodesic Equation}
\label{sec:geo}

\subsection{The Equations of Motion}

The dynamics of a test particle moving solely under gravity around a central mass in a static spacetime can be described using the geodesic formalism. Restricting motion to the equatorial plane $\theta=\pi/2$, the Lagrangian corresponding to the metric~(\ref{eq:metric}) is  
\begin{equation}
{\cal{L}}=\left[1- \Phi -\frac{r^2H^2}{c^2}\right]\dot{t}^2+\frac{2rH}{\sqrt{1-\Phi }}\dot{r}\dot{t}-\frac{\dot{r}^2}{1-\Phi} -r^2\dot{\phi}^2,
\end{equation}
 {where overdots denote derivatives with respect to the proper time $\tau$. The Euler–Lagrange equations for $x^{\mu}=\{t,r,\phi\}$ yield the corresponding geodesics. In the case of stationary, spherically symmetric spacetimes such as the Schwarzschild metric first integrals of the $t$ and $\phi$ equations can be found directly, while the $r$ equation is often replaced by the normalization condition: $g_{\mu\nu}\dot{x}^{\mu}\dot{x}^{\nu}=1$, (for timelike curves). This allows one to construct an energy-like equation for $r(\tau)$, from which $\ddot{r}$ may be obtained by differentiation. In contrast, the metric under consideration here depends on both $r$ and $t$, and thus the usual procedure cannot be directly applied. The radial geodesic equation retains an explicit $\ddot{r}$ term, which is essential for understanding the effective forces acting on the particle.}

 {The geodesic equation for the azimuthal angle reduces to $l = \dot{\phi} r^2$, where $l$ is the particle’s specific angular momentum. Substituting this into the radial equation and rearranging gives ~\cite{Nandra:2011ug,Nandra:2011ui}:}
\begin{align}
\ddot{r}=&\frac{l^2}{r^3}\left(1-\frac{3}{2}\Phi\right)-\frac{\Phi}{2}+rH^2 - \frac{2r^2 \dot{H}}{1-\Phi-r^2H^2}\dot{r}\dot{t}\nonumber\\
&+\frac{\dot{H} r S}{1-\Phi-r^2 H^2}\left(1+\frac{l^2}{r^2}\right)+\frac{r\dot{H}\dot{r}^2}{S\left(1-\Phi-r^2 H^2 / c^2\right)}.\label{eq:fullgeodesic}
\end{align}
 {Equation~(\ref{eq:fullgeodesic}) provides the exact form of $\ddot{r}$, valid in the full relativistic regime. In the weak-field limit assuming:}
\begin{equation}
r_s \equiv \frac{G M}{c^2} \ll r \ll D_H ,\quad \Phi \ll 1,     
\end{equation}
 {with $\dot{t}\approx 1$, $\dot{r}\approx 0$, we find the equation of motion to give:}
\begin{equation}  
\ddot{r} = \frac{l^2}{2r^3} -\frac{G M}{r^2} - \frac{(1 + 3w)}{2} H^2 r. 
\label{eq:local_expansion}
\end{equation}  
 {The term $(1 + 3w)H^2/2$ is often written as $ \equiv \ddot{a}/a$. An analytical solution for this equation of motion is studied in \cite{Baushev:2019mrg} as integral terms. Here we connect the spherical collapse model with modifications close to the velocity surface as a solution. }

\begin{figure}[t!]
\centering  
\includegraphics[width=0.4\textwidth]{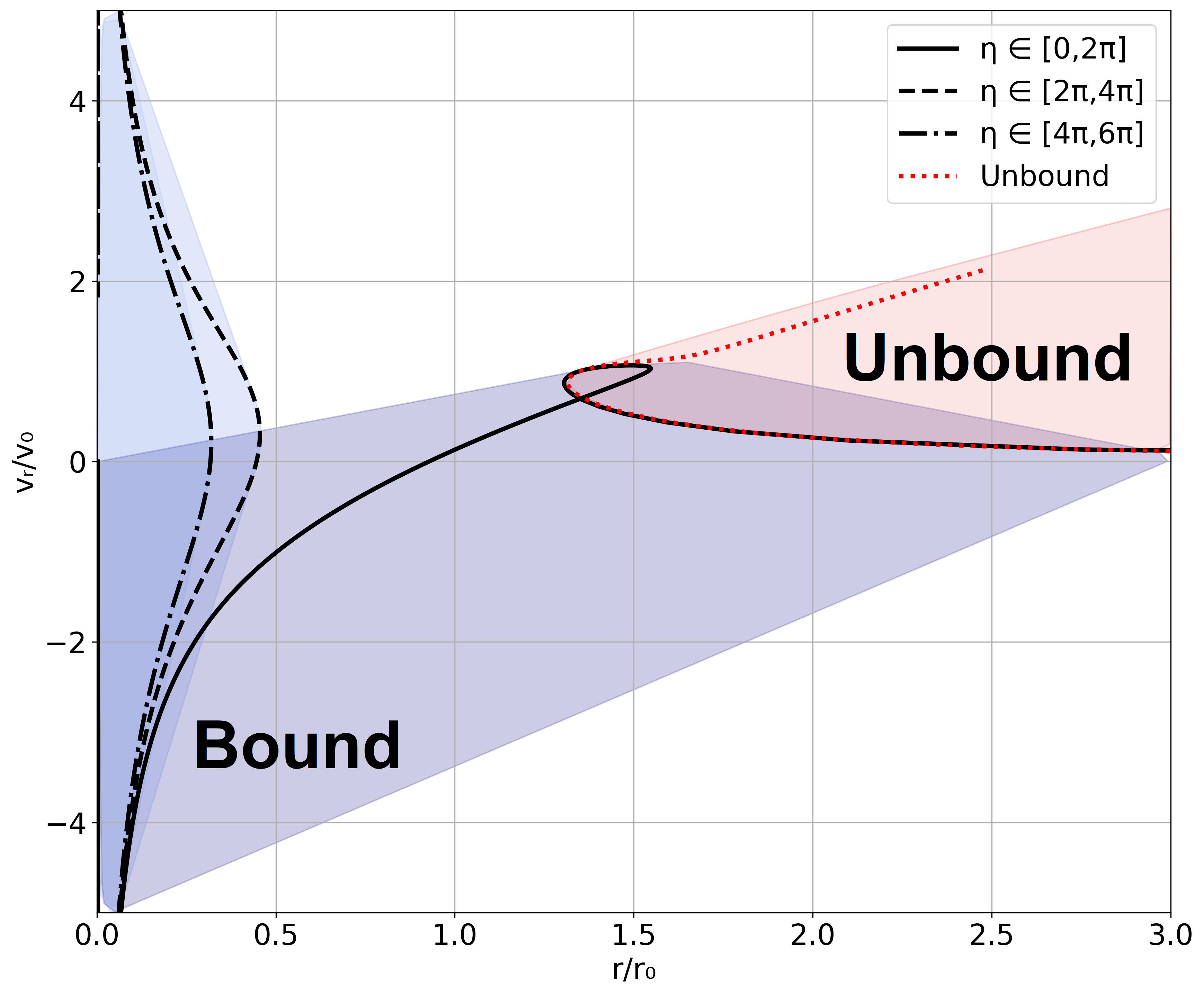} 
\caption{\it{The normalized velocity-distance relation from the analytical solution~(\ref{eq:sol_new_v_r}). The solution is scaled with $r_0$ and $r_0 /t_U$, where $t_U$ is the age of the Universe. The turnaround $r_0$ is related to the enclosed mass. The colored lines corresponds to $e = 1$. For different $e$ the area between the colored lines is also covered. The dashed line marks the asymptotic behavior for the unbound solution (which in the case of $H \neq 0$ corresponds to the Hubble flow).}}  
\label{fig:Newtonian_Hubble_flow}  
\end{figure} 

\subsection{Hubble Flow Around Virilized Structures}

\begin{figure*}[t!]  
    \centering  
\includegraphics[width=0.5\textwidth]{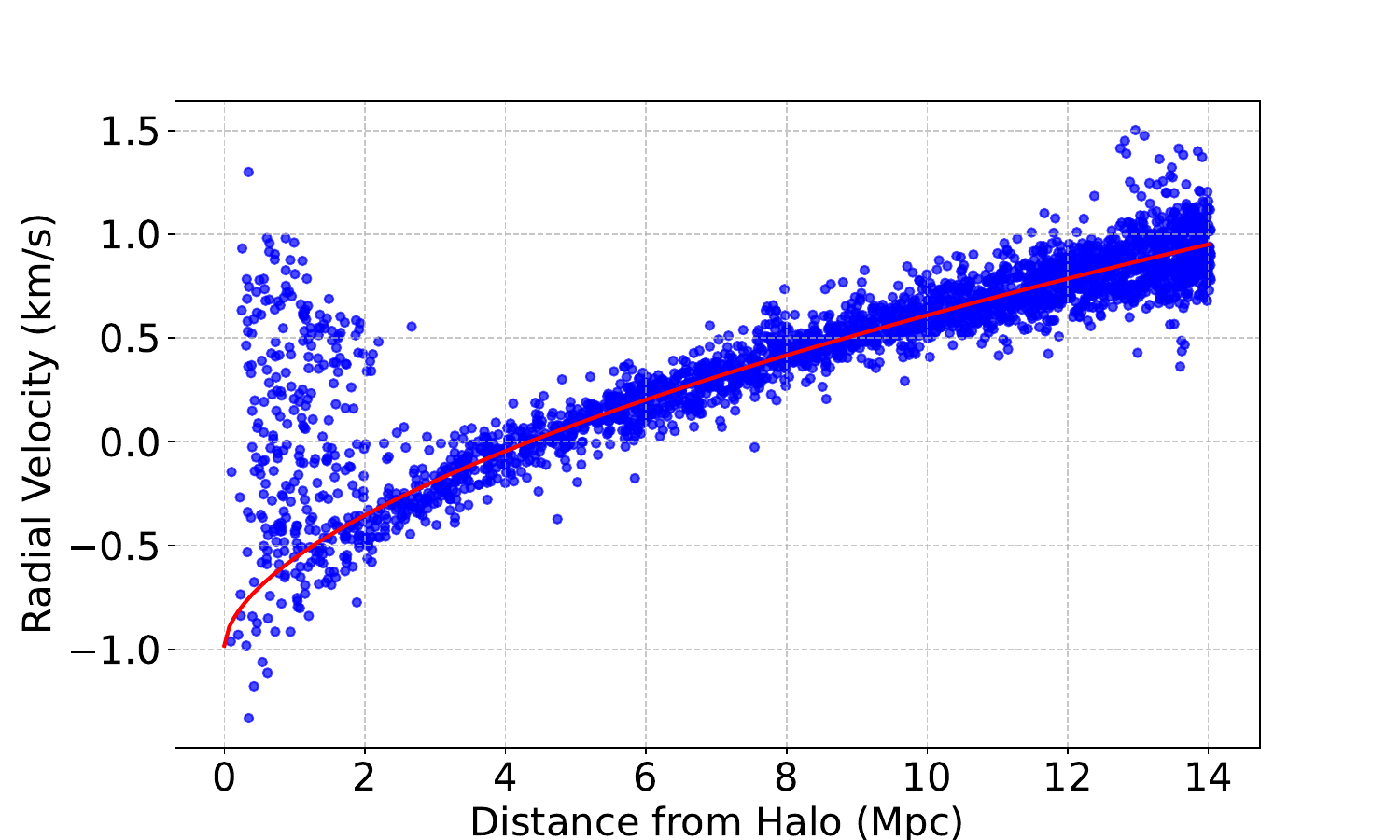} 
\includegraphics[width=0.47\textwidth]{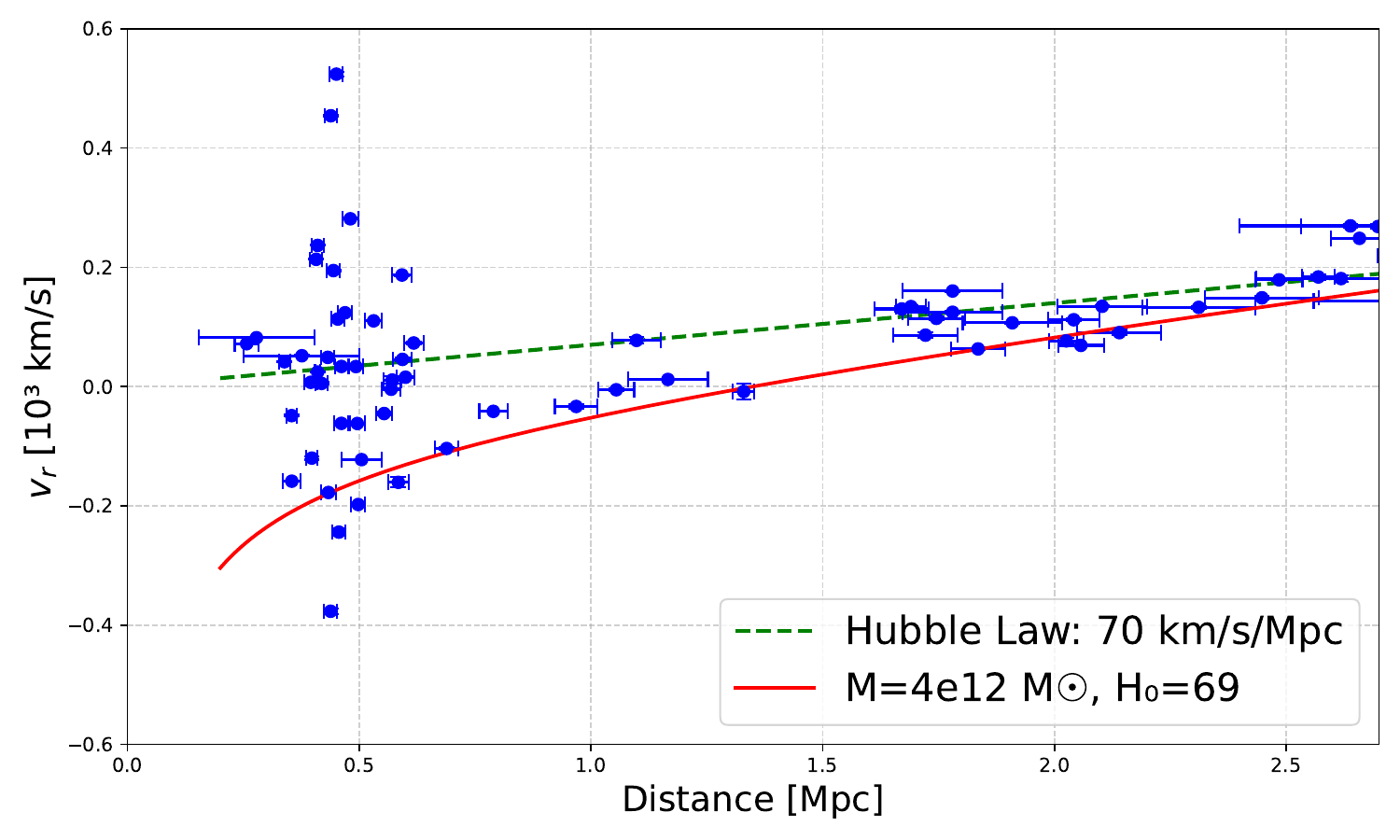}  
\caption{\it{\textbf{Left:} The velocity vs. distance for an isolated pair from the IllustrisTNG simulation. \textbf{Right:} The radial velocity of galaxies as a function of their distance from the halo CoM. A green dashed line illustrates the theoretical Hubble Law with a slope of $70\, \text{km/s/Mpc}$, representing the expected outward velocity for galaxies due to the universe's expansion.}}  
\label{fig:Hubble_flow}  
\end{figure*}

We can solve the motion analytically only for the Newtonian case, The canonical Lemaitre–Tolman model describes the dynamics of a spherically symmetric dust cloud under the influence of gravity, without the presence of a cosmological constant ($\Omega_\Lambda = 0$). The model provides analytical solutions for both bounded and unbounded systems, characterized by the eccentricity parameter $e$ and the phase parameter $\eta$. These solutions describe the evolution of the radial coordinate $r$, the radial velocity $v_r$, and the tangential velocity $v_t$ as functions of time $t$. For bounded systems, where $e \in [0,1]$ the radial coordinate $r$ and time $t$ are given by~\cite{Karachentsev:2008st}:
\begin{equation}
r = \frac{r_{\text{ta}}}{2} \left( 1 - e \cos \eta\right)\;, \quad t =\frac{t_{\text{ta}}}{\pi} \left( \eta - e \sin \eta \right)\;,
\label{eq:r_t_new}
\end{equation}
where $r_{\text{ta}}$ and $t_{\text{ta}}$ are the turnaround radius and turnaround time, respectively. The radial and tangential velocities are:
\begin{equation}
v_r =\frac{\pi}{2} \frac{ r_{\text{ta}}}{t_{\text{ta}}} \frac{e \sin \eta}{1 - e \cos \eta} \;, \quad v_t = \frac{\pi}{2} \frac{ r_{\text{ta}}}{t_{\text{ta}}} \frac{\sqrt{1-e^2}}{1 - e \cos \eta} \;.
\end{equation}
These equations describe the oscillatory motion of the system, where the radial coordinate reaches its maximum value at $\eta = \pi$. For the unbounded dwarf galaxies, the radial coordinate $r$ and time $t$ are expressed as:
\begin{equation}
r = \frac{r_{\text{ta}}}{2} \left( e \cosh \eta - 1\right)\;, \quad t =\frac{t_{\text{ta}}}{\pi} \left( e \sinh \eta - \eta \right)\;,
\end{equation}
with the corresponding velocities:
\begin{equation}
v_r =\frac{\pi}{2} \frac{ r_{\text{ta}}}{t_{\text{ta}}} \frac{e \sinh \eta}{e \cosh \eta - 1} \;, \quad v_t = \frac{\pi}{2} \frac{ r_{\text{ta}}}{t_{\text{ta}}} \frac{\sqrt{1-e^2}}{ e \cosh \eta - 1} \;.
\end{equation}
This solution describes a system that expands indefinitely, with the radial coordinate growing without bound as $\eta$ increases. The enclosed mass $M_{\text{TA}}$ is given by:
\begin{equation}
M_{\text{TA}} =  \frac{\pi^2 r_{\text{ta}}^3}{8 G t_{\text{ta}}^2},
\end{equation}
where $t_U$ is the age of the Universe. This expression is essentially a manifestation of Kepler's third law, relating the mass of the system to its dynamical timescale. For bounded systems, $\epsilon$ is negative, indicating a gravitationally bound state:
\begin{eqnarray}
r = r_0 \frac{1 - e \cos\eta}{(\eta - e\sin\eta)^{2/3}}\;, \quad
v_r = \frac{r_0}{t_U}  \frac{ e \sin\eta \left( \eta -e \sin\eta\right)^{1/3} }{1- e \cos\eta}, 
\label{eq:sol_new_v_r}
\end{eqnarray}
and the unbound solution:
\begin{eqnarray}
r = r_0 \frac{e \cosh\eta - 1}{(e \sinh\eta - \eta)^{2/3}}\;, 
\quad v_r = \frac{r_0}{t_U}e \sinh \eta\frac{\left( e \sinh \eta - \eta \right)^{1/3} }{e \cosh\eta - 1}.
\label{eq:sol_new_v_r_unbound}
\end{eqnarray}
Continuing from Fig.~(\ref{fig:Newtonian_Hubble_flow}), the dashed and dot-dashed lines within the bounded region $e \in [0,1]$ illustrate that dwarf galaxies experiencing stronger gravitational encounters (i.e., higher $\eta$ values) exhibit velocities corresponding to smaller radii. This trend indicates that such galaxies occupy dynamically tighter orbits due to energy dissipation from close interactions, confining them to the bound phase of the solution. In contrast, unbound trajectories ($e > 1$) at larger $r$ (shaded regions) reflect systems with insufficient binding energy for stable orbits. The concentration of dwarf galaxies at low $r$ underscores the role of environmental effects in shaping their kinematic profiles, aligning with expectations of Newtonian dynamics in high-density regimes.

\section{Expansion Domination}  
\label{sec:expan}
To evaluate the DE’s dominance, specific criteria are utilized. These parameters identify the regimes where DE’s repulsive influence overcomes gravitational attraction, governing the evolution of cosmic systems.

\subsection{Hubble Flow around Galaxy Groups and Clusters} 

\begin{table*}[t!]
\begin{tabular}{|l|c|c|c|l|}
\hline
System & $r_{ta}$ [Mpc] & $M_{vir} [M_{\odot}]$ & Ref taken & Others \\
\hline\hline
Virgo & $6.76 \pm 0.35$ & $(5.7 \pm 1.5)\times 10^{14}$ & \cite{Kim:2020gai} & \cite{Sorce:2016yok,Karachentsev:2010nw} \\
\hline
Coma & $6.67\pm0.08$ & $[1.04,2.77]\times 10^{15}$ & \cite{Benisty:2025tct} & \\
\hline
LG & $0.96 \pm 0.03$ & $\left(2.47\pm 0.15\right)\times 10^{12}$ & \cite{Makarov:2025} & \cite{Karachentsev:2008st,Penarrubia:2014oda} \\
\hline
Fornax & $[3.88,\,5.60]$ & $[1.30,\,3.93]\times 10^{14}$ & \cite{Nasonova:2011md} & \\
\hline
M81/M82 & $0.89 \pm 0.05$ & $(2.32 \pm 0.31)\times 10^{12}$ & \cite{karachentsev2002} & \\
\hline
CenA/M83 & $1.55 \pm 0.13$ & $(6.4 \pm 1.8)\times 10^{12}$ & \cite{Karachentsev:2006ww} & \\
\hline
\end{tabular}
\caption{\textit{ {The selected values for the fitted turnaround and enclosed mass from different references. }}}
\label{tab:data}
\end{table*}

The Hubble flow describes the motion of galaxies due to the expansion of the universe. On smaller scales, such as within galaxy groups or clusters, gravitational interactions cause deviations from this uniform expansion. \cite{bib:Sandage1986} introduced a method to estimate the mass of a system using the Hubble Flow. It is possible to jointly constrain the mass of the bound structure and $H_0$ from an observed velocity-distance diagram. This approach has been applied to galaxy groups in \cite{bib:Peirani2006,bib:Peirani2008,bib:Penarrubia2014,bib:Sorce2016,bib:DelPopolo2022}, and to galaxy clusters in \cite{bib:Kim2020,bib:Nasonova2011}. Beyond the spherical case, the domination of DE can be determined by the turnaround of galaxy groups and clusters, as discussed in~\cite{Kim:2020gai,Pavlidou:2013zha,Pavlidou:2014aia,Tanoglidis:2014lea}. For a discussion how to connect the line-of-sight velocities into the radial velocities, see Ref.~\cite{bib:Karachentsev2006,bib:Karachentsev2010,Wagner:2025wrp}.

 {The velocity distance relation for the Hubble flow line fit assumes the zero peculiar velocity at the big bang $v_{\text{pec}} = 0$ and initiate different test particles with initial radial velocities for today under the influence of effectively one body~\cite{Peirani:2005ti,Penarrubia:2014oda} and extended in~\cite{DelPopolo:2022sev}. Ref.~\cite{Peirani:2005ti,Penarrubia:2014oda} use different semi-analytical solutions for the local interplay with the Hubble flow. Ref.~ \cite{Peirani:2005ti} suggest the relation:}
\begin{equation}
v(r) = 1.377 H_0 r - 0.976 \frac{H_0}{r^n}\left(\frac{G M}{H_0^2}\right)^{(n+1)/3},
\label{eq:hubble_flow_1}
\end{equation}
with $n = 0.96$. Ref.~\cite{Penarrubia:2014oda} suggest a similar relation for a binary motion:
\begin{equation}  
v(r) = \left(1.2 + 0.31 \Omega_\Lambda \right) \, H_0 \, r - 1.1 \, \sqrt{\frac{G M }{r}},  
\label{eq:hubble_flow_2}  
\end{equation}  
 {where $\Omega_\Lambda$ is the DE rate. The red line in the figure shows the corresponding fit for the interplay branch.} 

 {In order to see the Hubble Flow behavior we test cosmological simulation together with observational data. The cosmological magnetohydrodynamic simulations analyzed in this work were drawn from the IllustrisTNG project \citep{bib:Vogelsberger2014,bib:Nelson2015}. We specifically used the TNG50~\citep{Nelson:2019jkf,Pillepich:2019bmb}. This simulation models a $\Lambda$CDM universe in a periodic box of side length $51.7~\mathrm{Mpc}$ with sufficient resolution to resolve galaxy-scale dynamics. The simulation adopts cosmological parameters consistent with Planck 2018 results~\citep{Planck:2018vyg}. To identify LG-like systems, we selected isolated galaxy pairs from TNG50 using the zero-velocity surface criteria from~\cite{bib:Sandage1986}. For a halo pair with a total mass $M$, the turnaround radius, $r_\mathrm{ta}$, is defined as the boundary at which inward peculiar velocities balance the Hubble expansion. Fig.~(\ref{fig:Hubble_flow}) shows an isolated halo from the IllustrisTNG simulations~\cite{Nelson:2018uso} with a mass of $\sim 10^{12} \, M_{\odot}$. The left panel illustrates the bounded and unbounded subhalos around the turnaround. We added the corresponding Hubble Flow fit in red line, based on the relation~(\ref{eq:hubble_flow_2}). The numerical fits shows a good agreement also on the simulation based analysis.}

\begin{figure*}[t!]  
    \centering  
\includegraphics[width=0.65\linewidth]{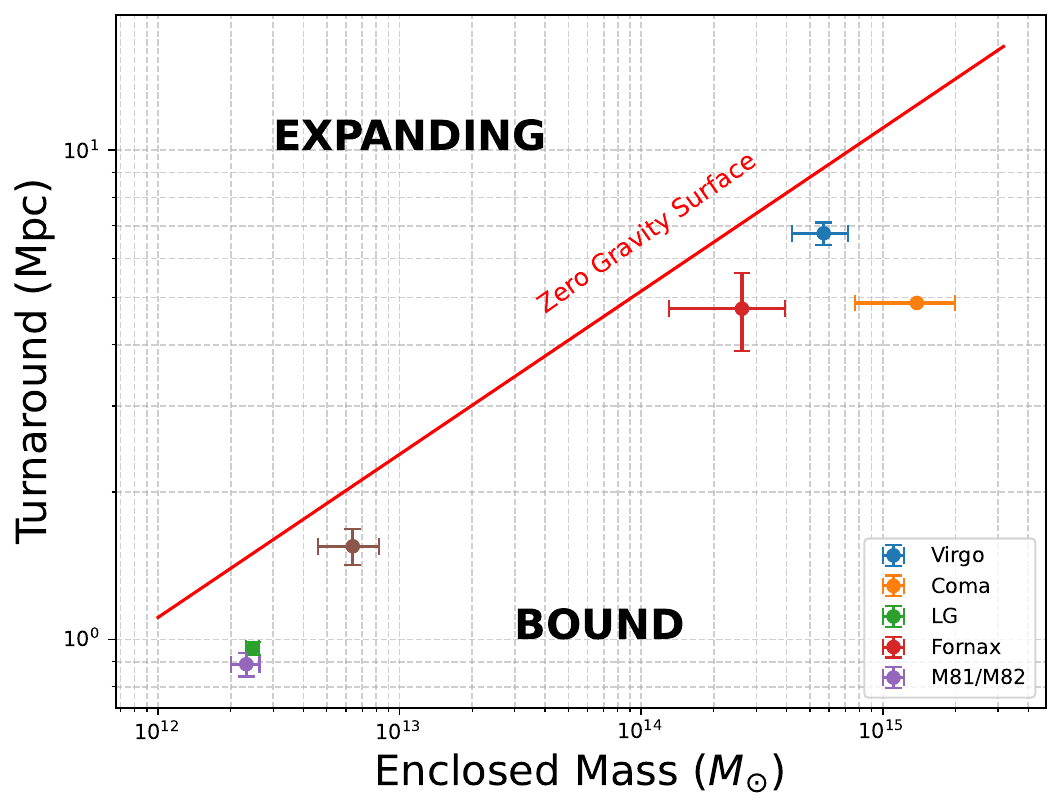}  
\caption{\textit{ {The enclosed mass versus turnaround radius (following~\cite{Pavlidou:2013zha}) for galaxy groups and clusters. The data points are taken from the references listed in Table under Fig.~\ref{fig:cur_phase_space}, including~\cite{Kim:2020gai,bib:Sorce2016,Benisty:2025tct,Makarov:2025,Karachentsev:2008st,Penarrubia:2014oda,Nasonova:2011md,karachentsev2002,Karachentsev:2006ww,DelPopolo:2021hkz}. The red curve indicates the \textit{Zero Gravity Surface}, defined by the balance condition $r_\Lambda$ which separates gravitationally bound systems from those dominated by cosmic acceleration.}}}  
    \label{fig:m_vs_r}  
\end{figure*}  

 {For comparison, we consider observational data on dwarf galaxies in the LG of galaxies taken from~\cite{2012AJ....144....4M,2020AJ....160..124M,2020RNAAS...4..229M,Tully:2022rbj}\footnote{\href{https://www.cadc-ccda.hia-iha.nrc-cnrc.gc.ca/en/community/nearby/}{LG and Nearby Dwarf Galaxies Link}}. We transform the velocities and the distances into the LG center of mass frame. Using the angular separation on the sky $\theta$ the physical distance $r_{\rm{gc}}$ between the galaxy pair is determined by:}
\begin{equation}
r_{\rm{gc}}^2 = r_{\rm{g}}^2 + r_{\rm{c}}^2 - 2 r_{\rm{c}} r_{\rm{g}} \cos\theta,
\end{equation}
 {where $r_{\rm{g}}$ and $r_{\rm{c}}$ are the distances of the galaxies from the observer. We transform the observed velocities and positions to the center-of-mass frame of the LG. Since the velocity components perpendicular to the line of sight are challenging to observe, the analysis is based only on the observed line-of-sight components. To calculate the radial infall velocities of the galaxies onto the LG center, assumptions about the system dynamics need to be added to estimate the impact of the unobserved velocity components \cite{bib:Karachentsev2006,Wagner:2025wrp}. We consider the \textbf{major infall model} that fits for a gravitationally bound systems. The model approximate the radial infall velocity by projecting the velocity difference between a galaxy and the centre of mass onto the line of sight of the galaxy to obtain:}
\begin{align}
v_{\mathrm{r,maj}} = \frac{v_j - v_\mathrm{c} \cos \theta_{\mathrm{c},j}}{{r}_j - {r}_\mathrm{c} \cos \theta_{\mathrm{c},j}} \;r_{gc}. 
\label{eq:v_maj}
\end{align}
 {It thereby assumes that that $v_{\perp,\mathrm{c}} = v_{\mathrm{t}} = 0$, in which $v_\mathrm{t}$ denotes the tangential velocity. The motion of galaxies within the LG can be described by modeling their velocity field, which provides insights into the system's dynamics. The analysis begins by correcting observed heliocentric velocities to a Galactocentric frame~\cite{bib:Karachentsev2006,Karachentsev:2008st}. The LG barycenter is assumed to lie along the line connecting the MW and M31. Its position is determined by their mass ratio $\gamma = M_{MW} / M_{Tot}$. Assuming minimal spread for the velocity dispersion around the center of mass, we get the value $\gamma = 0.55$ which is consistent with earlier estimations~\cite{Karachentsev:2008st,Makarov:2025}. The right panel of Fig.~(\ref{fig:Hubble_flow}) shows the radial velocity vs. distance from the center of mass of the LG. The bounded area, which includes the MW, M31, and their dwarf galaxies, shows a spread up to the turnaround radius. Beyond this, we observe the local interaction with the local Hubble expansion. }

\begin{figure*}[t!]  
\centering  
\includegraphics[width=0.8\linewidth]{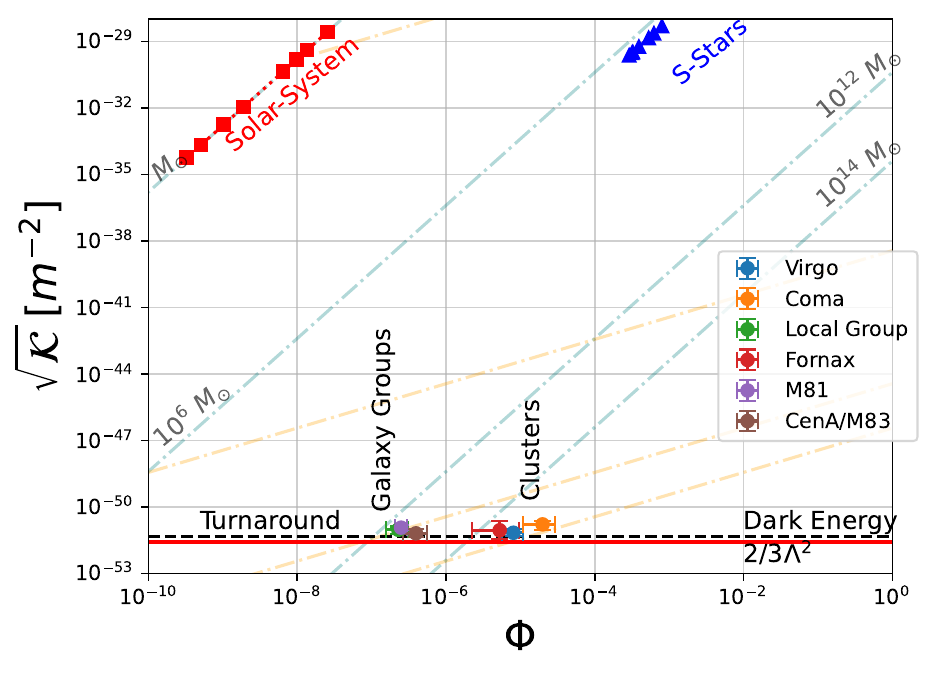} 
\caption{\textit{The curvature phase space for the potential $\Phi$ versus the scalar $\sqrt{\mathcal{K}}$ in a de-Sitter $w = -1$ universe. The Solar System and S-stars around the galactic center are positioned much higher than galaxy groups and clusters, which lie at the transition to local expansion, indicating the dominance of $\Lambda$ in larger structures. The table below gives the characteristic turnaround radii $r_{ta}$ and masses ($M$) of galaxy groups and clusters. Another Ref. for the turnaround and enclosed mass is taken from~\cite{Kim:2020gai,bib:Sorce2016,Benisty:2025tct,Makarov:2025,Karachentsev:2008st,Penarrubia:2014oda,Nasonova:2011md,karachentsev2002,Karachentsev:2006ww,DelPopolo:2021hkz}}}
    \label{fig:cur_phase_space} 
\end{figure*}

 {The maximal turnaround is also related to expansion rate, as found by~\cite{Pavlidou:2014aia}. Asymptotically, the term $\ddot{a}/a$ in Eq.~\eqref{eq:local_expansion} approaches}
\begin{equation}
\lim_{t\rightarrow\infty} \frac{\ddot{a}}{a} = \Omega_\Lambda H_0^2,
\end{equation} 
 {which is the value for a de Sitter universe. This result highlights that in the far future, the dynamics of the expansion is dominated entirely by the cosmological constant, erasing the influence of matter and radiation. Consequently, any bound system is eventually subject to the repulsive effect of DE, setting a natural limit to the largest scale on which gravity can counterbalance the accelerated expansion. Put differently, there exists a maximum radius beyond which the gravitational attraction of a mass $M$ can no longer hold a test particle against the outward push of DE. At this critical radius, the two effects are exactly in balance. The maximal turnaround obeys the condition $\ddot{r} = 0$ leading to:}
\begin{equation}
r_{\text{ta}} \leq r_{\Lambda} \equiv \sqrt[3]{\frac{G M}{\Omega_\Lambda H_0^2}}.
\label{eq:max_turnaournd}
\end{equation}
 {This distance represents the maximal limit for a test particle within this potential. Ref.~\cite{Pavlidou:2014aia} discusses the systems and their turnaround radii, showing that, for bounded galaxy groups and clusters, this is the maximal turnaround. For example, the LG has a maximum turnaround of $1.5\, \text{Mpc}$ (from Eq.~\eqref{eq:max_turnaournd} with $M = 3\cdot 10^{12} \, M_{\odot}$), while \cite{Peirani:2005ti,Penarrubia:2014oda} reports $r_{\text{ta}} \approx 1\, \text{Mpc}$.}

The curvature scalar $\mathcal{K}$ and dimensionless parameter $\kappa$ for galaxy groups and clusters are written as:  
\begin{align}  
\mathcal{K} = \mathcal{C} (1 + 2\kappa),  \quad
\kappa = \left( \frac{r}{r_\Lambda} \right)^3.  
\end{align}  
In the context of curvature scalars, the maximal turnaround modifies the Kretschmann scalar to:
\begin{equation}
\left[\mathcal{K}\right]_{\text{ta}} = 3\left[\mathcal{K}\right]_{\text{Cos}},
\end{equation}
where the cosmological Kretschmann scalar is:
\begin{equation}
\left[\mathcal{K}\right]_{\text{Cos}} \equiv 12/D_H^2.
\end{equation}
Here, spacetime curvature is dominated by the background expansion, as the contribution from concentrated mass becomes negligible. If the curvature nears the cosmological value, the expansion dominates motion, as seen for dwarf galaxies outside the turnaround. For the bounded region, the Kretschmann scalar is much higher and Newtonian dominated, indicating negligible $\Lambda$ effects.

To study the dynamics of galaxy groups and clusters, we analyze systems such as the CenA/M83, LG, M81/M82 groups, and the Virgo, Fornax, and Coma clusters~\cite{Karachentsev:2013cva,Tully:2022rbj,Kim:2020gai,Nasonova:2011md,Makarov:2025,Karachentsev:2006ww,Peirani:2005ti,Peirani:2008qs,DelPopolo:2021hkz,DelPopolo:2022sev}. For each system, we adopt the characteristic turnaround radii $r_{ta}$ and total enclosed masses $M_{\rm vir}$ summarized in table ~(\ref{tab:data}) and in Fig.~(\ref{fig:m_vs_r}). Galaxy groups typically have masses of $\sim 10^{12}\,M_{\odot}$, while clusters occupy the $\sim 10^{14}\,M_{\odot}$ or higher mass range. Using these parameters, we compute the corresponding curvature scalars, allowing a direct comparison between local gravitational binding and the influence of cosmic expansion. All of the bounded structures are under the zero-gravity line.

The curvature parameters gives another phase space to compare.  {Since some curvature scalar vanishes at different cases, we compare $\Phi$ and the $\mathcal{K}$, as~\cite{Baker:2014zba}.} As illustrated in Fig.~(\ref{fig:cur_phase_space}), all analyzed systems align closely with the $\left[\mathcal{K}\right]_{ta}$ line. This indicates that galaxies near the turnaround radius occupy a transitional regime between the dominance of the Hubble flow and the onset of local gravitational collapse, highlighting how DE begins to influence the dynamics of larger structures while smaller, tightly bound systems remain largely decoupled from cosmic expansion.

\subsection{Spherical Density} 
 {Another case that our formalism can be formulated with is the spherical density case. The curvature scalar for spherical density $\rho_m$ in an expanding $\rho_\Lambda $ reads: } 
\begin{eqnarray}  
\mathcal{R} = \frac{32 \pi G \rho_\Lambda}{c^2}, \quad \mathcal{C} = \frac{ 256 \pi^2 G^2 \rho_m}{c^4} , \quad \mathcal{K} = \mathcal{C} \left( 1 + \Delta_c\right),    
\end{eqnarray}  
where parameter $\kappa$ (from Eq.~\ref{eq:curv_scalar}) changes is the known $\Delta_c$, which is the ratio between the densities:
\begin{equation}
\kappa = \Delta_c \equiv \frac{\rho_m}{\rho_\Lambda},
\end{equation}
with $\rho_\Lambda \equiv 8 \pi G/\Lambda c^2$. The parameter emerges in the Kretschmann scalar and quantifies the interplay between the local density of the object and the global cosmological density. This highlights the connection between local gravitational effects and the large-scale influence of DE.  

This transition is formalized in the Spherical Collapse Model~\cite{1972ApJ...176....1G,1980lssu.book.....P}, which outlines the growth of a small initial density perturbation. The overdensity gradually detaches from the Hubble flow, expands to a maximum turnaround radius and begins to collapse. Collisionless dynamics arrest the collapse, leading to virialization at a final radius which is half of the turnaround radius. The critical overdensity $\Delta_c$ quantifies the contrast between the virialized density and the background density at collapse time in a matter-dominated universe. This yields: $\Delta_c \approx 18\pi^2 \approx 178$, implying that virialized halos are approximately 178 times denser than their cosmological surroundings. $\Lambda$CDM cosmology predicts marginally lower values due to the influence of DE~\cite{Pace:2017qxv}.

 {To determine the impact of the local expansion for galaxy groups and clusters we have to take into account objects nearly the turnaround. However, from the spherical collapse picture it gives characteristics for the extension of the bound region of a density.} The spherical density framework demonstrates how curvature scalars incorporate the competing influences of local matter density and the uniform background set by DE. This same reasoning can be extended from single, spherically symmetric overdensities to dynamical multi-body systems. In particular, binary motion provides another context in which the interplay between local gravitational binding and cosmic expansion manifests itself in the curvature invariants.

\subsection{Binary Motion}  
 {The last case we investigate under the curvature perspective is in the context of binary motion. In the case of binary objects, the curvature scalars for binary systems can be expressed as~\cite{Benisty:2023vbz,Benisty:2023clf,Strigari:2025nqa},:}  
\begin{eqnarray}  
\mathcal{C} = \frac{12 \omega_{\text{kep}}^2}{c^4},  
\quad \mathcal{K} =  \frac{12 \omega_{\text{kep}}^2 + 24 H^4}{c^4} = \mathcal{C} (1 + 2\kappa),  
\end{eqnarray}  
where $\omega_{\text{kep}}$ is the Keplerian orbital frequency. The ratio $\kappa$ naturally emerges from $\mathcal{K}$, highlighting the interplay between orbital dynamics and DE: 
\begin{equation}
\kappa \equiv \left(\frac{T_{\text{Kep}}}{T_H}\right)^2,
\end{equation}
where $T_{\text{Kep}} = 2\pi \sqrt{r^3/GM}$ is the Keplerian orbital period, and
\begin{equation}
T_H \equiv \frac{2\pi}{H_0} \approx 27.6\, \text{Gyr}
\end{equation}
is the Hubble time, calculated from the measured Planck values~\citep{Aghanim:2018eyx}. Whereas the spherical density analysis emphasizes virialization and collapse in overdense regions, the binary framework emphasizes the long-term stability of bound orbits against the slow influence of cosmic acceleration. To probe the effect of $\Lambda$, one must identify systems with orbital periods close to $T_H$. Notably, $T_H$ is much larger than the age of the Universe ($\sim 13.7\, \text{Gyr}$), indicating that DE's influence on binary systems is subtle and becomes significant only over cosmological timescales.

The LG provides the longest orbital period among bounded binary systems discussed in~\cite{Benisty:2023vbz}. Since the orbital period of the LG is about $\approx 0.62 T_H$, the upper bound on $\Lambda$ is expected to be tighter than those derived from the solar system or the S2 star. Ref.~\cite{Benisty:2023clf} finds the upper bound on $\Lambda$ to be 5 times larger than the Planck value \cite{Planck:2018vyg}. Ref.~\cite{Benisty:2023clf} further shows that two main characteristics determine the effect of $\Lambda$: the accuracy of the measurements and the period of the system. Although the solar system provides more accurate measurements than the LG, the upper limit on $\Lambda$ derived from the solar system is about $10^{-46} \, \text{m}^{-2}$, which is approximately 10 orders of magnitude larger than the actual value. In contrast, the LG's longer orbital period results in a tighter upper limit, demonstrating the importance of system timescales in constraining DE.

\section{Discussion}  
\label{sec:dis}
This research explores the dominance of DE across various astrophysical systems by analyzing curvature scalars. We unify the description of gravitational effects from small-scale systems, such as the Solar System, to large-scale structures like galaxy clusters, using the McV spacetime as a theoretical framework. By examining the interplay between local gravitational dynamics and cosmological expansion, we identify regions where DE dominates using the scalar invariant from General Relativity.

 {From the McV spacetime, the dimensionless parameter  $\kappa$ emerges, which marks a critical transition point across all systems studied. Of particular significance is the emergence of a universal dimensionless parameter, $\kappa$, from the curvature scalars of the McVittie spacetime. Despite originating from different physical contexts—be it the dynamics of galaxies, the collapse of overdensities, or the orbits of binaries—this parameter consistently delineates the frontier where DE's repulsive influence begins to dominate over local gravitational attraction. For galaxy groups and clusters, $\kappa \equiv (r/r_\Lambda)^3$ (from Eq.~\ref{eq:curv_scalar}). The condition $\kappa = 1$ defines the maximal turnaround radius $r_\Lambda$ (Eq.~\ref{eq:max_turnaournd}), the point of zero acceleration where a test particle is precisely balanced between the inward gravitational pull and the outward push of cosmic acceleration. This represents the absolute limit of the gravitational influence of a structure. Our analysis shows that the observed turnaround radii of local structures align closely with this $\kappa \sim 1$ transition, as evidenced by their position on the curvature phase space (Fig.~\ref{fig:cur_phase_space}) and their distribution relative to the zero-gravity surface (Fig.~\ref{fig:m_vs_r}). This indicates that these massive structures exist at the very threshold of being unbound by DE.}

 {In the spherical collapse model, the same parameter $\kappa$ represents the density contrast $\Delta_c = \rho_m / \rho_\Lambda$ (Eq.~\ref{eq:curv_scalar}). Here, $\kappa = 1$ ($\rho_m = \rho_\Lambda$) signifies the scale where local matter density equals the homogeneous DE density. While virialization itself occurs in the deep potential wells where $\kappa = \Delta_c \gg 1$, the $\kappa = 1$ threshold defines the larger cosmological boundary within which the collapse process is embedded. The presence of DE modifies the critical overdensity for collapse from its Einstein-de Sitter value \cite{Pace:2017qxv}, demonstrating that $\Lambda$ influences structure formation well before the $\kappa = 1$ boundary is reached.}

 {For binary systems, $\kappa$ takes the form $(T_{\text{Kep}}/T_H)^2$. The condition $\kappa = 1$ corresponds to an orbital period equal to the Hubble time ($T_{\text{Kep}} = T_H$), representing a fundamental upper limit on the orbital period of any gravitationally bound system in an accelerating universe. No known binary system approaches this limit; the longest-period systems, such as the Milky Way-Andromeda binary ($T_{\text{Kep}} \approx 0.62 T_H$, $\kappa \approx 0.38$), remain in the gravity-dominated regime ($\kappa < 1$) but already exhibit measurable effects from $\Lambda$ \cite{Benisty:2023clf}.}

 {This multi-scale analysis reveals a consistent narrative: while the specific definition of $\kappa$ is context-dependent, the $\kappa \sim 1$ transition universally marks the scale where a system's dynamics are no longer dictated solely by local gravity but are fundamentally reshaped by the cosmic acceleration driven by DE. Our formalism, based on the curvature invariants of the McVittie spacetime, provides a powerful and consistent geometric language to identify and characterize this transition across the vast range of scales in the cosmos.}

The curvature scalars serve as key indicators of this transition, emerging naturally from fundamental theory and providing a unified framework for understanding the influence of DE. Fig.~(\ref{fig:cur_phase_space}) illustrates the curvature phase space for the potential $\Phi$ versus the scalar $\sqrt{\mathcal{K}}$ in a de Sitter universe. The Solar System and S-stars are positioned much higher in the phase space, reflecting strong local gravitational effects, while galaxy groups and clusters lie at the transition to local expansion, where DE dominates. This contrast highlights the scale-dependent influence of DE: smaller systems remain dominated by local gravity, whereas larger structures exhibit curvature scalars that approach their cosmological values, indicating the dominance of DE. By comparing systems of varying sizes, we have shown that DE dominates in larger structures, such as galaxy groups and clusters, where the curvature scalars align with cosmological predictions. This phase-space analysis provides a clear and unified framework for understanding the interplay between local gravitational dynamics and the large-scale expansion of the universe driven by DE.

 {The cases studied in this research are based on idealized configurations. To extend the analysis to more realistic environments, such as galaxy groups and clusters, we employ N-body simulations. As shown in Fig.~\ref{fig:Hubble_flow}, the theoretical predictions align well with both the simulation results and the observational data of the LG. The McV framework provides an exact solution of Einstein’s equations for a single point-like mass embedded in an expanding FLRW background. This construction relies on spherical symmetry and the absence of energy exchange between the local source and the cosmological fluid. These assumptions, however, restrict its applicability: once multiple localized masses are considered, isotropy and spherical symmetry are broken, and no straightforward generalization of McV exists. Consequently, McV should be regarded as an idealized model, valid only for isolated systems. To overcome these limitations, we turn to cosmological simulations such as IllustrisTNG, which capture the complexity of realistic cosmic environments. The agreement between the simulated Hubble flow around large-scale structures and the predictions of the McV picture confirms the robustness of our analytical framework despite its idealizations. This approach not only clarifies the conditions under which DE dominates local dynamics but also offers a geometric perspective for interpreting observations of cosmic expansion.}

\acknowledgments
DB thanks Salvatore Capozziello and Jenny Wagner for useful discussions and suggestions. DB is supported by a Minerva Fellowship of the Minerva Stiftung Gesellschaft fuer die Forschung mbH.

\bibliographystyle{apsrev4-1}
\bibliography{ref.bib}

\end{document}